\documentstyle[preprint,aps,epsf,prb]{revtex} 
\begin{document} 

\tightenlines

 
\title{\bf Isotopic effects of hydrogen adsorption in carbon nanotubes
}   
\author{M.C. Gordillo, J. Boronat, and J. Casulleras}  
\address{ 
Departament de F\'{\i}sica i Enginyeria Nuclear, Campus Nord B4-B5, \\  
Universitat Polit\`ecnica de Catalunya. E-08034 Barcelona, Spain}  
\date{\today} 
 
\maketitle 
\begin{abstract} 
We  present diffusion Monte Carlo calculations of D$_2$ adsorbed inside
a narrow carbon nanotube. The 1D D$_2$ equation of state is reported, and
the one-dimensional character of the adsorbed D$_2$ is analyzed.
The isotopic dependence of the constitutive properties of the quantum fluid are
studied by comparing D$_2$ and H$_2$. Quantum effects due to their
different masses are observed  both in the energetic and the structural properties.
The influence of the interatomic potential in  one-dimensional systems
is also studied by comparing the properties of D$_2$ and $^4$He which have
nearly the same mass but a sizeably different potential. The physics of
molecular hydrogen adsorbed in the interstitial channels of a bundle of
nanotubes is analyzed by means of both a diffusion Monte Carlo 
calculation and an approximate mean field method.
\end{abstract} 
 

\section{INTRODUCTION} 

Since their discovery by Iijima in 1991,\cite{iji} carbon nanotubes and
their
fascinating properties have attracted the attention of both theoretical and
experimental physicists.\cite{SDD98,DDES98,aja}
Nanotubes are basically long cylinders, their walls being formed from one 
(single-walled carbon nanotubes, SWCN) or several (multiple-walled 
ones, MWCN) graphite sheets. One of its more important features is 
its extremely narrow width, with diameters in the nanometer scale, ranging from 
$\sim$ 7 to 40 \AA, compared to its length, thousands of times larger. 
The enormous length to diameter ratio and the narrowness of the tube make nanotubes 
excellent systems to hold inside nearly perfect one-dimensional fluids.
Moreover, as a consequence of its narrowness, the adsorption energy of an
atom or molecule inside a carbon nanotube is several times greater than 
on planar graphite. This feature ensures that any particle of adequate size 
will be swallowed inside carbon nanotubes, provided that the half-fullerene 
caps at their ends are removed. Its use as a possible solution to the packing 
of H$_2$ in fuel cells has been discussed in the literature. \cite{dil,dar,stan1,rze,wan}

Apart from its technological interest, the quasi-one dimensionality of
these carbon tubes is itself an
interesting property from a purely theoretical point of view. In the same way
that a sheet of graphite provides an almost perfectly 2D experimental 
environment, the small transversal room available to the particles
adsorbed in a narrow carbon nanotube creates a proper setup to extract 
experimental information of quasi-one-dimensional systems. Additionally, if the
temperature is low enough, quantum effects become important, and an 
experimental realization of a quasi-one-dimensional quantum fluid is realized.
A first step in that direction has been recently given
by Teizel {\em et al},\cite{tei} who unambiguously observed the quasi-one dimensional 
behaviour of $^4$He adsorbed inside SWCN bundles. The helium intake
could be directed either to the inside of the nanotubes, to the interstitial
channels between the several tubes constituting a bundle, or 
to the external surface of the nanotube bundle. The occupation
probability in the three regions depends on the geometry of the bundle,
on the {\em size} of the atom/molecule to be adsorbed and, finally, on the
interatomic potential between the carbon atom and each particular species.
\cite{cole}
In all cases, an almost perfect alignment of the atoms is granted.
Encouraged by this experimental accessibility, some theoretical work about the 
adsorption of quantum liquids inside nanotubes has been carried out. 
The most studied case is $^4$He, which has been considered both in 1D and 
inside a (5,5)
nanotube. In the limit of zero temperature, both 1D and  quasi-one-dimensional 
$^4$He are self-bound systems, with a binding energy ranging the milikelvin
scale.\cite{kro,gor,moroni} 

Molecular hydrogen inside nanotubes is also a very appealing system. In fact,
the confinement of H$_2$ in SWCN, in the limit of zero temperature, 
has recently deserved theoretical work.
Its mass is approximately half the $^4$He one, but the hydrogen-hydrogen 
interatomic potential has a potential well three times deeper than the 
helium-helium one. This prevents the existence of a liquid phase at zero
temperature in 
both 2D and 3D, but diffusion Monte Carlo (DMC) calculations\cite{yo} 
have shown
that it is not enough to preclude it in 1D. This result is
due to the reduced number of neighbours of each atom. In fact, this liquid state could 
be produced also in higher dimensions when the number of molecules surrounding
a particular one is artificially reduced, as in small clusters\cite{sin} (3D)
and in surfaces doped with the right kind of impurities\cite{gor2} (2D). 

In the present work, the DMC method is used to study the influence of 
both the interparticle potential and mass 
on the thermodynamic behavior of the isotopes of molecular hydrogen, 
especially H$_2$ and D$_2$, adsorbed inside a carbon nanotube.
Since the electronic structure of molecular deuterium and hydrogen 
is the same, the D$_2$-D$_2$ interparticle potential is
identical to the H$_2$-H$_2$ one. This is equally true 
for the particle-tube interactions. Moreover, the mass of the 
D$_2$ molecule is very similar to that of a $^4$He atom. Thus, by
comparing the deuterium results with those for helium, 
the effect of the respective potential wells can be inferred. On the
other hand, the influence of the 
zero-point energy in the thermodynamic behavior of a quasi-one 
dimensional array can be drawn from the comparison between the deuterium and
hydrogen results. In the last part of the paper, we present results 
for the equation
of state of H$_2$, D$_2$ , and $^4$He when they are adsorbed in the
interstitial channels of a nanotube bundle. 
The influence of neighbouring arrays in a particular one, as a function of the 
interchannel separation, is studied using the DMC equation of state of 
purely 1D systems. The
attractive Van der Waals tails between the arrays increase the binding
energy of the adsorbed liquid in a quantity which decreases with the
interchannel separation. 
Using the DMC method, it is verified that a {\em mean field} approximation, 
as suggested in Ref. \onlinecite{cole2}, is very accurate at realistic separations.

The outline of the paper is as follows. In the next section, we briefly
introduce the DMC algorithm used to study both the fluid adsorbed in a
nanotube and the purely one-dimensional model. In Sec. III, we compare
results of the equation of state and spatial
structure for the two isotopes, H$_2$ and D$_2$. Predictions for the
equations of state of both isotopes, when the adsorbtion is in the
interstitial channels, are also reported. Finally, Sec. IV comprises the
summary and the main conclusions of our work.

\section{METHOD} 

The diffusion Monte Carlo method, \cite{les,boro4he} is a theoretical tool 
that has proved its accuracy in a large variety of systems and physical scales. 
It solves in a stochastic way the Schr\"odinger equation, providing results
for the ground state which are exact for boson systems like the present
one. The $N$-body Schr\"odinger equation is solved in imaginary time 
after the introduction of importance sampling, a standard technique 
to reduce the variance and make the algorithm operative. Explicitely,
\begin{eqnarray}
-\frac{\partial f({\bf R},t)}{\partial t} & = & -D {\mbox{\boldmath $\nabla$}}^2 
f({\bf R},t) + D {\mbox{\boldmath $\nabla$}} ({\bf F} \, f({\bf R},t)) \\
\nonumber
 & & + (E_{\rm L}({\bf R}) -E) f({\bf R},t) \ ,
\label{dmc1}
\end{eqnarray}
with  $f({\bf R},t) = \psi({\bf R})\, \Psi({\bf R},t)$,
where $\psi({\bf R})$ denotes  a trial wave function used for importance sampling
and $\Psi({\bf R},t)$ the imaginary-time-dependent wave function.
In Eq.  (\ref{dmc1}), $D = \hbar^2/2m$,
$E_{\rm L}({\bf R}) = \psi({\bf R})^{-1} H \psi({\bf R})$ is the local energy, 
and  ${\bf F}({\bf R}) = 2\, \psi({\bf R})^{-1}
{\mbox{\boldmath $\nabla$}} \psi({\bf R})$ is the so-called drift force
which guides the diffusion movement to regions where $\psi({\bf R})$ is
large.

DMC calculations have been carried out for two different situations:
{\it i}) a pure 1D array of D$_2$ molecules, and {\it ii}) D$_2$ filling
a tube of radius $R$ = 3.42 \AA, corresponding to the
(5,5) armchair nanotube in the standard nomenclature.\cite{tipus} 
This structure has been chosen because it is one of
the narrowest nanotubes experimentally obtained, and therefore it is 
expected to be close to an ideal 1D arrangement. 

In the present calculation, we have considered the D$_2$ molecules
interacting
via the Silvera and Goldman (SG) pair potential.\cite{sil1} This semiempirical 
interaction
is the same that has been used in Ref. \onlinecite{yo} in the study of H$_2$ in tubes, and
has become the standard approach in molecular hydrogen and molecular
deuterium calculations. 
The SG interaction is a spherically averaged model, even though the
pair of molecules involved are ellipsoids. However, the eccentricity of the
ellipsoids is so small ($r$ = 0.94, in the H$_2$ case) that it produces results of remarkably 
accuracy, even at low temperatures. 
For the D$_2$-tube interaction we have used
the Stan and Cole (SC)\cite{stan1,stan2} H$_2$-tube interaction. 
 This is a cylindrically averaged potential, i.e., 
the potential felt by a molecule inside the tube is a function only of its
distance to the center of the tube.  
We have verified that the differences between the results
obtained with this averaged interaction and the ones derived by explicitely
summing up
all the C-D$_2$ interactions inside the tube are negligible. 
The SC potential depends only on three parameters: $\sigma$, $\epsilon$, 
and the radius of the tube. For the (5,5) tube considered, $R = 3.42$ \AA,
and  the parameters $\sigma$ and $\epsilon$  correspond to the Lennard-Jones
C-H$_2$ interaction ($\sigma=2.97$ \AA \ and $\epsilon=42.8$ K).\cite{stan2} 
In Fig. 1, the H$_2$-tube SC potential is displayed and compared
to the $^4$He potential inside the same cylinder.
 The two curves reflect the differences between the pair 
($\sigma$, $\epsilon$) for $^4$He and H$_2$; the larger values of
$\epsilon$ and $\sigma$ in the case of H$_2$ originates a deeper well and
a shorter range, respectively.

The trial wave function $\psi$ used in the 1D system is

\begin{equation}
\Psi^{\rm 1D}({\bf R}) = \Psi_{\rm J}({\bf R})  \ ,
\label{psi1d}
\end{equation}
with $\Psi_{\rm J}({\bf R}) = \prod_{i<j} \exp \left( -0.5 \, \left
(b/\,r_{ij} \right)^5
\right)$ a Jastrow wave function with a McMillan two-body correlation factor.   

For the simulation of D$_2$ inside the tube, an additional
one-body factor is introduced in order to avoid the hard core of the
tube-molecule interaction,
\begin{equation}
\Psi^{\rm T}({\bf R}) = \Psi_{\rm J}({\bf R}) \Psi_{\rm c}({\bf R})   \ ,
\label{psitub}
\end{equation}
with $\Psi_{\rm c}({\bf R}) = \prod_i^N \exp(-c \, r_i^2)$, $r_i$ being the
radial distance of the particle to the center of the cylinder.

The parameters $b$ and $c$ entering in $\Psi_{\rm J}$ and $\Psi_{\rm c}$,
respectively, have been determined by means of a variational Monte Carlo
(VMC) optimization. Near the equilibrium density we find the optimal
value $b = 3.996$ \AA, gradually increasing with the linear density $\lambda$ (at 
$\lambda = 0.358$ \AA$^{-1}$, $b =4.026$ \AA).
The parameter $c$, inversely proportional to the width of the gaussian
defined in $\Psi_{\rm c}$, has been  fixed to $c= 6.392$ \AA$^{-2}$ 
due to its negligible dependence on the density. 
 It is worth mentioning that these parameters are different from
the ones used for H$_2$ in the same setup\cite{yo} ($b=3.759$ \AA, $c=4.908$
\AA$^{-2}$).

\section{RESULTS} 
 
\subsection{Low-density regime} 

In order to study the behavior of the several isotopes of molecular
hydrogen inside nanotubes, the dependence of the adsorption energy on
the mass species constitutes a first relevant point. Since the 
interparticle and tube-molecule potentials do not distinguish between 
isotopes, their different masses play the central role. At low densities,
the dominant effect is due to the binding energy of single independent 
molecules.  From our DMC calculations the binding energy of a single 
molecule inside a carbon tube turns out to be, for
the (5,5) tube,
$-1539.87 \pm 0.11$ K for H$_2$ , and $-1605.23 \pm 0.01$ K and $-1624.37 \pm
0.01$ K for 
D$_2$ and T$_2$, respectively. The increase in the binding energy with the mass
comes from the combination of two features: a decrease in the kinetic
energy, mainly due to a direct effect of the mass ($m_{{\rm H}_2}/m _{{\rm
D}_2} \simeq 1/2$, $m_{{\rm D}_2}/m _{{\rm T}_2} \simeq 2/3$), and a
simultaneous increase of the potential energy.
Although those
energies correspond to the ground state of single molecules at 0 K,
they constitute a very good estimation in the limit of infinite dilution
at nonzero temperatures.
From the above binding energies one can extract information on the
selectivity in the adsorption inside the nanotube.
Following Ref. \onlinecite{sie}, the selectivity of isotope 2 with 
respect to isotope 1 
can be defined by the quotient
$S=(x_1/x_2)/(y_1/y_2)$ with $x_i$ ($y_i$) the nanotube (bulk) mole
fractions. It has been proved that in the limit of zero pressure the selectivity
$S_0$ is very well approximated by
\begin{equation}
S_0 = \frac {m_2}{m_1} \ \exp \left( - \frac{E_{1}-E_{2}}{T} \right) \ ,
\label{select}
\end{equation}
where $E_i$ is the binding energy of isotope $i$. 
Considering $T=20$ K, as in Ref. \onlinecite{sie}, we obtain 
$S_0({\rm T}_2 / {\rm H}_2 ) = 22.8$ and $S_0({\rm T}_2 / {\rm D}_2 ) = 1.7$
for the (5,5) tube.  The selectivity is especially high in
the case T$_2$/H$_2$ due to the sizeable difference in binding energies
between the two isotopes, $E_{{\rm T}_2} - E_{{\rm H}_2} = -84.5$ K. That
large selectivity, which is a purely quantum effect, has been proposed in
Ref. \onlinecite{sie} to achieve an efficient isotopic sieving.

\subsection{One- and quasi-one-dimensional systems}

The ground-state properties of an array of hydrogen molecules, in a 
one- and in a  quasi-one-dimensional environments, have also been 
studied using the DMC method. 
In a previous work,\cite{yo} we had studied  the case of H$_2$:
the equilibrium state in 1D corresponds to a liquid phase 
with a  density $\lambda_0 = 0.2191 \pm 0.0004$ \AA$^{-1}$ and
an energy per particle $(E/N)_0 = -4.834 \pm 0.007$ K.
A first approach to determine the equation of state of 1D D$_2$ 
is based on the use of the  average correlation approximation (ACA).
This approximation, which has been widely used in the $^3$He-$^4$He
isotopic mixture in the limit of zero $^3$He concentration,\cite{aca1,aca2}
consists in the
present case in evaluating the expectation value of the Hamiltonian for
D$_2$ using the {\em exact} wave function of the other isotope (H$_2$). 
Thus, the energy per particle of D$_2$ in ACA is given by
\begin{equation}
   \left( \frac{E}{N} \right)_{{\rm D}_2} = 
   \left( \frac{E}{N} \right)_{{\rm H}_2} + \left( \frac{m_{{\rm H}_2}}
   {m_{{\rm D}_2}} - 1 \right) \left( \frac{T}{N} \right)_{{\rm H}_2} \ ,  
\label{acaener}
\end{equation}
where $(T/N)_{{\rm H}_2}$ is the kinetic energy of the H$_2$ molecule.

ACA corresponds, obviously, to a variational estimate and thus the 
D$_2$ energies
obtained from Eq. (\ref{acaener}) are upper bounds to the exact values.
Using the DMC kinetic energies of 1D H$_2$ reported in Table I, and  the total
energies published in Ref. \onlinecite{yo}, we have calculated the ACA 
equation of state for D$_2$. The results obtained are shown in Fig. 2 
and compared with the exact 
solution derived from a DMC calculation of 1D D$_2$. Both 
equations of state (lines) correspond to a
least squares fit to the data in the usual form
\begin{equation}
\frac{E}{N} = \left( \frac{E}{N} \right)_0 +  A \left( \frac{ 
\lambda - \lambda_0 }{\lambda_0} \right)^2 +
B \left( \frac{ \lambda - \lambda_0 }{\lambda_0} \right)^3 \ .
\label{eqestat}
\end{equation}
The  parameters  $(E/N)_0$, $\lambda_0$, $A$, and $B$ are given in Table II. 
The upper bounds provided by ACA  appear clearly reflected in Fig. 2, its
{\em quality} being quite reasonable taking into account its
straightforward estimation (\ref{acaener}).
In fact, the discrepancies between ACA and the exact results are less than 
2 \% for the equilibrium density value, and around 3 \% 
for the binding energies. ACA can also be used to estimate the equation of
state of the T$_2$ fluid, a system that we have not studied here using the 
explicit DMC calculation. For 1D T$_2$, ACA predicts  $\lambda_0 = 
0.2501$ \AA$^{-1}$ and $(E/N)_0 = -12.677$ K.

On the other hand, the equation of state of D$_2$ inside a tube can 
also be estimated by ACA, since the tube-molecule SC potential does not 
depend on the particular isotope. Using the H$_2$ equation of state 
(Ref. \onlinecite{yo}) and the H$_2$ kinetic energies
reported in Table I, the ACA results compare with the corresponding DMC
ones with an accuracy similar to the 1D case. DMC energy results for both 
1D D$_2$ and D$_2$ adsorbed in the (5,5) nanotube
are displayed in Fig. 3.
In order to show the two equations of state with the same energy
scale, we have subtracted to the tube results the adsorption 
energy of a single molecule.
In Fig. 3, the curves are  polynomial fits (\ref{eqestat}) to the DMC data,
the optimal parameters for the tube being reported in Table II. It 
can be seen that the equilibrium density for 1D D$_2$ and D$_2$ adsorbed 
in the (5,5) nanotube are almost identical. This is also true for
the location of the spinodal points of D$_2$, 
$\lambda_{\rm s}^{\rm 1D} = 0.230 \pm
0.001$ \AA$^{-1}$ and $\lambda_{\rm s}^{\rm T} = 0.232 \pm 0.001$
\AA$^{-1}$,
which can be derived from the data contained in Table II.

The binding energies, in the respective equilibrium points ($(E/N)_0$), 
are slightly different: the additional transverse degree of freedom 
only amounts to an increase of 0.091 K. This increase in the binding energy 
is nearly a factor two smaller than the one drawn from the DMC calculations 
for H$_2$ (0.172 K). In relative terms, the increase of the binding energy 
is only a 0.85 \% for D$_2$ versus a 3.5 \% for H$_2$.
Therefore, one can conclude that the effects of the additional degree of 
freedom of the D$_2$ molecules in the radial direction inside the nanotube, 
which account for the enhancement of the binding energy, are reduced by 
the greater mass of the D$_2$ molecule with respect
to the H$_2$ one. As a matter of comparison, it is illustrative to compare
the effects observed in D$_2$ with the ones previously studied in $^4$He
using the same methodology and geometry. It is worth noticing that the
masses of D$_2$ and $^4$He are nearly the same whereas the interatomic
potentials are sizeably different. The DMC results show that the latter effect
is completely determinant: in  $^4$He the relative difference mentioned 
above is 90 \%, two orders of magnitude 
larger than in D$_2$. Another minor effect that contributes to the 
one-dimensionality of molecular deuterium adsorbed inside the tube, 
is the larger hard-core size of the 
C-D$_2$ interaction ($\sigma_{{\rm C}-{\rm D}_2} = 2.97$ \AA), versus the
C-He one
($\sigma_{{\rm C}-{\rm He}} = 2.74$ \AA). 
The mass versus potential effects can also be seen in the value of the
equilibrium density. Inside the tube, $\lambda_0$ goes from $0.079 \pm 0.003$
\AA$^{-1}$ in $^4$He, to $0.2200 \pm 0.0006$ \AA$^{-1}$ in H$_2$, to reach
$0.2473 \pm 0.0002$ \AA$^{-1}$ in D$_2$. That sequence clearly indicates
that the main 
influence in $\lambda_0$ comes from the potential, since the isotopic change 
varies the
location of the energy minimum less than 15 \%. Most probably, the features 
observed in these systems have a more general character and one can guess
that if the well of the interatomic potential is increased 
and/or the mass of the particle adsorbed inside a tube is enlarged, the effect
would be an increase in the $\lambda_0$ value. 

In Fig. 4, the density dependence of the pressure for both  
1D D$_2$ and D$_2$ inside the tube 
is reported from the equilibrium density up to 0.30 \AA$^{-1}$.  
For the sake of comparison, the same results for H$_2$ are
also plotted.  In both H$_2$ and D$_2$, the pressures for the
1D systems (solid lines) are greater than the ones for the tubes (dashed
lines) at the same $\lambda$, the difference being quite similar for 
the two isotopes. 
For the same geometry and at equal density, the pressure of H$_2$ is always 
larger, a fact that is basically due to its slightly smaller
equilibrium density.

Complementary and useful information on the system can be obtained from the
microscopic study of the spatial structure of the molecules in the array.
In Fig. 5 the radial distribution functions, $g_z(r)$, along the $z$ axis, 
are shown.  They correspond to the quantum fluids adsorbed in the tube at
their respective equilibrium densities $\lambda_0$. 
The 1D counterparts are nearly identical and are not displayed. 
Being the denser of the three systems, D$_2$ exhibits accordingly the
most pronounced oscillations in the $g_z(r)$ function.
The shift in the positions of the maxima for the two molecular isotopes
arises basically from the difference in their respective $\lambda_0$'s.  
A comparison of the radial distributions of H$_2$ and D$_2$ at the same 
linear density, not
shown for simplicity, indicates so. In the $^4$He case, the much lower 
equilibrium density, direct consequence of the different potential,
explains the smoothness of the  $g_z(r)$ obtained.
    
The radial densities inside the (5,5) tube have been also studied.
In Fig. 6,  the radial densities for $^4$He, H$_2$, and D$_2$
for the same linear density $\lambda$ = 0.245 \AA$^{-1}$ are shown. 
The trends shown in the figure are common to all densities studied:
the particle 
with the largest  mass (D$_2$) is the one which spends more time in regions 
closer to the center of the tube, i.e., D$_2$ in the tube is the closest to a
one-dimensional system. The change in the mass  and in the
interatomic potential  work in the same direction: the radial
densities of H$_2$ and $^4$He are quite similar. Both curves show a
decrease in the radial localization and larger fluctuations
in the transverse direction.

\subsection{Adsorbtion in a bundle of nanotubes}

The absorption of different species inside the interstitial channels of 
bundles, formed when several nanotubes
are staked together, has been theoretically predicted\cite{cole} and 
experimentally observed.\cite{tei}  The staking of the nanotubes allows
for the formation of an hexagonal network of quasi-one dimensional channels
separated by  distances  which depend on the diameter of the nanotubes 
themselves. For instance, when the nanotube ropes are made out of (5,5)
constituents, the  distance between two adjacent tubes is 10.26 \AA, 
whereas the closest interstitial channels are located $\sim 5.9$ \AA \ apart.      
The interstitial channels are very narrow, with effective diameters oscillating
between 3 and 3.5 \AA. \cite{tersoff} This size is considerably smaller than
the diameter (6.8 \AA) of the (5,5) tube considered in the previous
subsection. Therefore, it is plausible to assume that if 
molecular hydrogen or deuterium is introduced inside one of this channels, 
its behavior would have even a more one dimensional character  than in
the inner part of a single tube. 

Considering the previous arguments, a quantum fluid  adsorbed in the intersites
can be well modelled by  hexagonal arrangements of purely 1D systems,
mutually separated by the real interchannel distance in the bundle. To a
large extent, the additional consideration of the interactions between the
adsorbed atoms/molecules and the carbon atoms of the nanotube walls would
simply introduce a constant shift in the energy scale. 
The influence of 
neighbouring channels in the binding energy of $^4$He adsorbed
in one of them has been estimated in Ref. \onlinecite{cole2} considering a {\em mean
field} approximation. In this approach, the neighbouring channels are seen
as uniform arrays that only contribute to increase the potential energy in
the form
\begin{equation}
\Delta_{\rm mf} = \frac{\lambda}{2} \int_{- \infty}^{\infty}  d x 
\ V \left(\sqrt{x^2 + d^2}\right) \ ,
\label{mfield}
\end{equation}
where  $d$ is the distance between the reference
channel and  one of its neighbours. The total energy correction is obtained by
summing up the contribution of the nearest neighbours, the next-nearest
neighbours, and then on  up to a desired accuracy. 
From its definition (\ref{mfield}) it is clear that the energy per particle 
in this approximation, 
\begin{equation}
E/N = (E/N)_{\rm 1D} + \Delta_{\rm mf} \ ,  
\label{enermf}
\end{equation}
is an upper bound to the exact energy. 

In order to test  the {\em mean field} approximation (\ref{enermf}),
we have performed full DMC calculations of a system formed by an hexagonal
array of 1D H$_2$ channels.
Since the distances between channels depend
on the diameter of the tubes, the calculations have been made in the 
more challenging case
for Eq. (\ref{enermf}), a bundle of (5,5)
nanotubes. That tube  is one of the narrowest experimentally
obtained, and therefore the  interchannel distance is one of the 
smallest in nature. 
Actually, that is an unrealistic case  since the
corresponding interstitial channels are  too narrow to adsorb H$_2$.
The DMC energies are displayed in Fig. 7, and compared with the approximated
estimations (\ref{enermf}). 
As it is evident in the figure,  both results are nearly identical, at
least in the density range considered. The conclusion is that 
correlations between  hydrogen molecules located in different channels 
suppose a negligible contribution to the energy, 
even in this case where the channels  are very close. In Fig.
7, there is also reported the equation of state of 1D H$_2$. As one can
see, the increase of the binding energy of the system, when the array is
immersed in a bundle, is as large as a factor two.

Having checked the validity of the {\em mean field} approach (\ref{enermf})
in an exigent case, one can use that approximation to study the equation of
state of molecular hydrogen and helium adsorbed in other more appropriate 
bundles. It is
worth mentioning that the number of neighbours to be included in
$\Delta_{\rm mf}$ has to be larger than ten, otherwise the binding 
energy  is underestimated in a  5 \%. Besides the (5,5) bundle, with an 
interchannel
distance $a=5.92$ \AA, we have analyzed the following cases:  
(6,6),  $a = 6.71$ \AA; (8,8),  $a  = 8.27$ \AA;
(10,10),   $a=9.83$ \AA. The fluids that have been analyzed are $^4$He,
D$_2$, and H$_2$. In Fig. 8, the differences between the equilibrium density
in the bundles and in the purely 1D system are shown for the three cases.
There is a clear difference (an order of magnitude) between the 
increase of $\lambda_0$, when $^4$He
is adsorbed in the intersites, and the density shift 
experimented by the two hydrogen isotopes. In H$_2$ and D$_2$ the change in
density is small but one can distinguish a slightly larger effect in the
lighter isotope. In all cases, the density difference decreases when the
interchannel separation increases. However, H$_2$ and D$_2$ have already
reached the 1D
equilibrium density in the (10,10) geometry whereas $^4$He still shows a
sizeable effect.  

The increase of the binding energies with respect to the one-dimensional
systems are reported in Fig. 9. The different points correspond to the same
bundles reported in Fig. 8. In absolute terms, that increase is much larger
in molecular hydrogen than in $^4$He, the effect being slightly larger in
D$_2$ than in H$_2$ for all the bundles. The energy difference between
H$_2$ and D$_2$ is very small and decreases when the interchannel distance
increases. It is clear, from the {\em mean field} correction
(\ref{mfield}),
that the difference in energy between the two isotopes arises from their
different densities since the interatomic potential is the same. In $^4$He
the shift in energy is comparatively small but, taking into account that
the 1D $^4$He binding energy\cite{gor} is only $(E/N)_0=-0.0036$ K, 
one concludes that the
bundle effects are much larger in $^4$He than in molecular hydrogen.

\section{Summary and Conclusions}

In this work, we have studied the zero-temperature equation of state of 
molecular deuterium, and compared the results with those previously
obtained\cite{gor} for
$^4$He and molecular hydrogen.\cite{yo} The results show that the 
different masses of
the two isotopes do influence both the energetic and structural properties.
As a general trend, D$_2$ appears as a slightly denser liquid, a fact that
is observed in its equation of state and also in the radial distribution 
functions. The
comparison between D$_2$ and $^4$He, two {\em particles} with the same mass
but with very different interatomic potentials, has shed light on the
influence of the potential in one-dimensional systems. Concerning the mass and
potential changes, the DMC results show unambiguously that the potential
effects are manifestly dominant. Bundles of carbon nanotubes have also been
studied by modelling them as hexagonal one-dimensional arrays. The
explicit DMC calculation of the system has allowed to check the accuracy of
a previously suggested {\em mean field} approximation.\cite{cole2} 
Once verified the validity of that approximation, we have analyzed the effects
of the neighbouring channels on the 1D equation of state. Our results
show that those effects are much more relevant in $^4$He than in molecular
hydrogen.

In the tube and bundle
calculations, we have assumed that the interparticle potential is not
modified by the existence of nearby carbon atoms. Recently, 
Cole and coworkers\cite{tri1,tri2} have estimated the effective correction in the
two-body potential due to three-body interactions between two particles and
the carbon atoms of the nanotube wall. This three-body interaction has been
modelled by the well-known  triple-dipole interaction derived by Axilrod and
Teller,\cite{axil} a potential energy contribution that is mostly repulsive.
In the case of molecular hydrogen in the interstitial channels, the well 
of the  effective Silvera and Goldman potential is 
reduced by approximately a factor two. This unexpected and big effect would
modify some of the results presented in our work. Nevertheless, it is
well-known that, in spite of the fact that 
Axilrod-Teller  is the dominant contribution to
$V_3$, at short interparticle distances an attractive force
emerges. This short-ranged  three-body interaction, known as exchange
interaction,\cite{bruch} is due to the influence of a third particle 
in the charge densities of two interacting atoms. Due to the short
interparticle distances H$_2$-C-H$_2$ in the channel, that attractive term
would partially cancel the repulsion introduced by Axilrod-Teller. We have
preferred to ignore the three-body corrections\cite{com1} until a better 
knowledge of
the nonadditive terms of this particular system is achieved.

We plan, in the near future, to introduce in the simulation 
the possibility of the swelling of the channels
when molecular hydrogen is adsorbed. The aim would be to test, by a
microscopic calculation, the recent theoretical result\cite{swell2} that  
H$_2$ forces the nanotube bundles to slightly spread upon adsorption. 
We hope our calculations, and the ones carried out by other theoretical
groups, can stimulate new experiments on molecular hydrogen adsorption in
carbon nanotubes and bundles, especially at low-temperature where 
quantum effects become macroscopic.

\acknowledgments
The authors gratefully acknowledge Prof. M. W. Cole for stimulating
discussions.
This research has been partially supported by DGESIC (Spain) Grant N$^0$
PB98-0922, and DGR (Catalunya) Grant N$^0$
1999SGR-00146. M. C. G. thanks the Spanish Ministry of Education and Culture
(MEC) for a postgraduate contract. We also acknowledge the supercomputer 
facilities
provided by the CEPBA.

\begin{table}
\begin{tabular}{ccccc}
        &  \multicolumn{2}{c}{1D}  & \multicolumn{2}{c}{Tube} \\
$\lambda$ (\AA$^{-1})$ &  H$_2$  &  D$_2$  &  H$_2$  &  D$_2$  \\ \tableline
0.225    & 9.36 $\pm$ 0.03   & 5.63 $\pm$ 0.02
         & 125.5 $\pm$ 0.2   & 80.5 $\pm$ 0.1   \\
0.230    & 10.20 $\pm$ 0.03  & 6.03 $\pm$ 0.02
         & 125.8 $\pm$ 0.2   & 80.9 $\pm$ 0.1   \\
0.235    & 11.32 $\pm$ 0.02  & 6.57 $\pm$ 0.02
         & 127.1 $\pm$ 0.2   & 81.2 $\pm$ 0.1    \\
0.239    & 12.30 $\pm$ 0.03  & 7.28 $\pm$ 0.02
         & 127.8 $\pm$ 0.2   &  81.8 $\pm$ 0.1    \\
0.259    & 18.27 $\pm$ 0.05  & 10.98 $\pm$ 0.03
         & 133.6 $\pm$ 0.2   & 85.4 $\pm$ 0.1     \\
\end{tabular}
\caption{Kinetic energies of H$_2$ and D$_2$ in 1D and inside a (5,5)
carbon nanotube.} 
\end{table}

\begin{table}
\squeezetable
\begin{tabular}{lccc}
          & ACA &  1D D$_2$  & D$_2$ in a tube \\ \tableline
$\lambda_0$ (\AA$^{-1})$ & 0.2419 $\pm$ 0.0007  & 0.2457 $\pm$ 0.0003 
& 0.2473 $\pm$ 0.0002 \\
$(E/N)_0$ (K)               & -10.282 $\pm$ 0.008   
& -10.622 $\pm$ 0.016 & -1615.94 $\pm$ 0.015 \\
$A$   (K)              & 154 $\pm$  3  & 2.0 10$^2$  
$\pm$ 1.0 10$^1$ & 2.13 10$^2$ $\pm$ 1.0 10$^1$ \\
$B$   (K)               & 4.5 10$^2$ $\pm$ 70  
& 9.6 10$^2$  $\pm$ 1.2 10$^2$ & 1.10 10$^3$ $\pm$ 1.1 10$^2$ \\
\end{tabular}
\caption{Parameters of the equation of state of D$_2$ (Eq.
\protect\ref{eqestat}).}
\end{table}

\begin{figure}
\begin{center}
\epsfxsize=6cm  \epsfbox{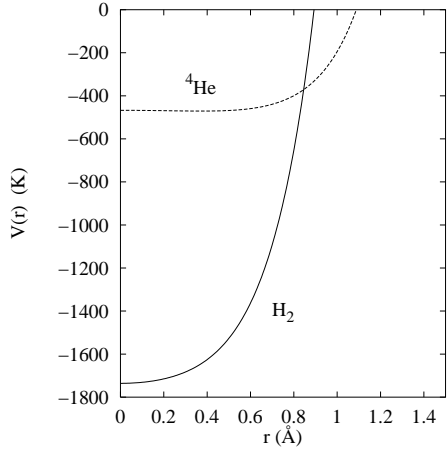}
\end{center}
\caption{SC tube-molecule/atom potentials inside a (5,5) carbon nanotube.}
\end{figure} 


\begin{figure}
\begin{center}
\epsfxsize=6cm  \epsfbox{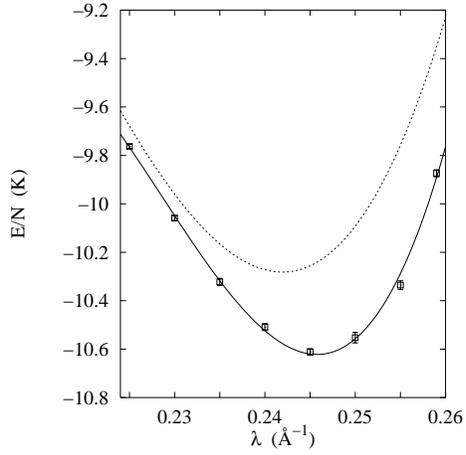}
\end{center}
\caption{Equation of state of  1D D$_2$. The symbols are the DMC energies
and the solid line a polynomial fit to them (\protect\ref{eqestat}). The
dashed line corresponds to the energies estimated using ACA.}
\end{figure} 

\pagebreak

\begin{figure}
\begin{center}
\epsfxsize=6cm  \epsfbox{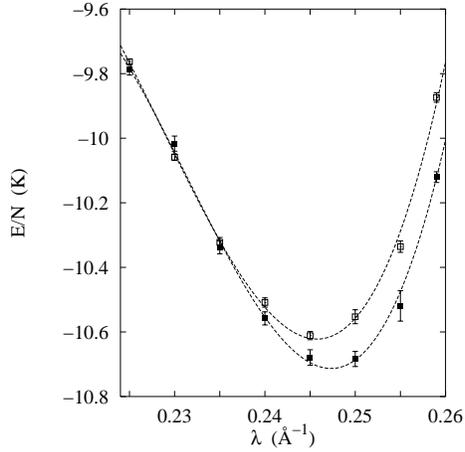}
\end{center}
\caption{Comparison between the equation of state of  1D D$_2$ and D$_2$
adsorbed in the nanotube. Filled squares correspond to the tube results;
open squares, to the 1D ones. The lines are polynomial fits to the DMC
data. To better compare both results, we have subtracted to the tube
energies the binding energy of a single molecule.} 
\end{figure}


\begin{figure}
\begin{center}
\epsfxsize=6cm  \epsfbox{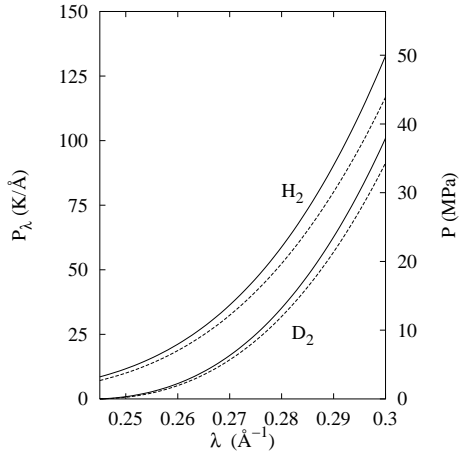}
\end{center}
\caption{Pressure of D$_2$ and H$_2$ as a function of the linear density
$\lambda$. Solid and dashed lines  correspond to the 1D and nanotube
systems, respectively. The left (right) scale is the 1D (nanotube)
pressure.} 
\end{figure}


\begin{figure}
\begin{center}
\epsfxsize=6cm  \epsfbox{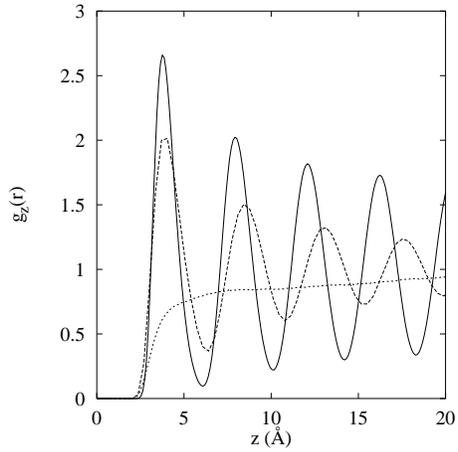}
\end{center}
\caption{Two-body distribution function in the nanotube system, and 
in the $z$ direction. Solid,
dashed, and dotted lines correspond to D$_2$, H$_2$, and $^4$He,
respectively. All curves are calculated at their respective equilibrium
densities $\lambda_0$.}
\end{figure}


\begin{figure}
\begin{center}
\epsfxsize=6cm  \epsfbox{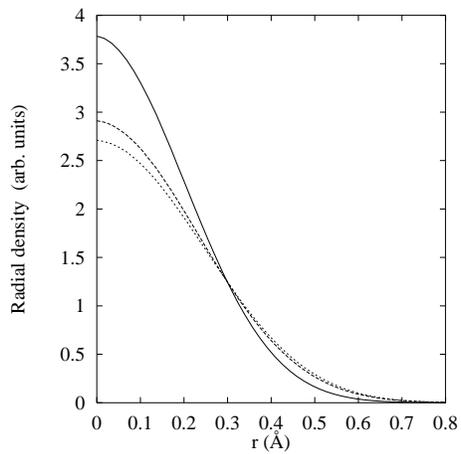}
\end{center}
\caption{Radial density of D$_2$ (solid line), H$_2$ (dashed line), and
$^4$He (dotted line) inside the (5,5) nanotube.}
\end{figure}

\pagebreak

\begin{figure}
\begin{center}
\epsfxsize=6cm  \epsfbox{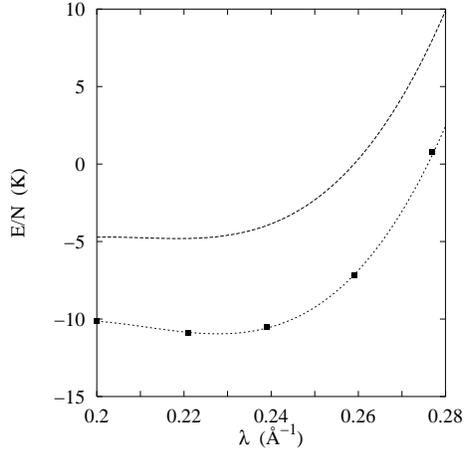}
\end{center}
\caption{Equation of state of D$_2$ in a bundle of (5,5) nanotubes. The
symbols are the DMC results, and the dotted line corresponds to the mean
field approximation. The dashed line is the 1D result.}
\end{figure}


\begin{figure}
\begin{center}
\epsfxsize=6cm  \epsfbox{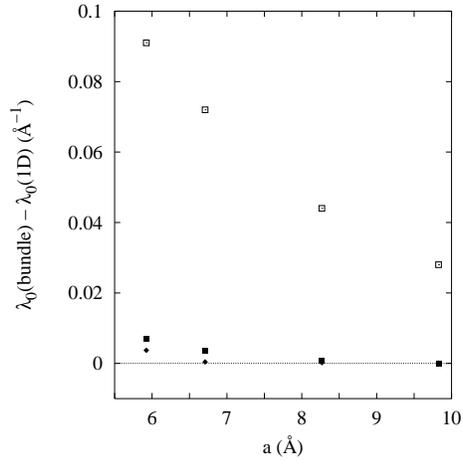}
\end{center}
\caption{Difference between the equilibrium densities, in a bundle and in a 1D
array, as a function of the cell parameter $a$. Diamonds, filled squares,
and open squares correspond to D$_2$, H$_2$, and $^4$He, respectively.}  
\end{figure}


\begin{figure}
\begin{center}
\epsfxsize=6cm  \epsfbox{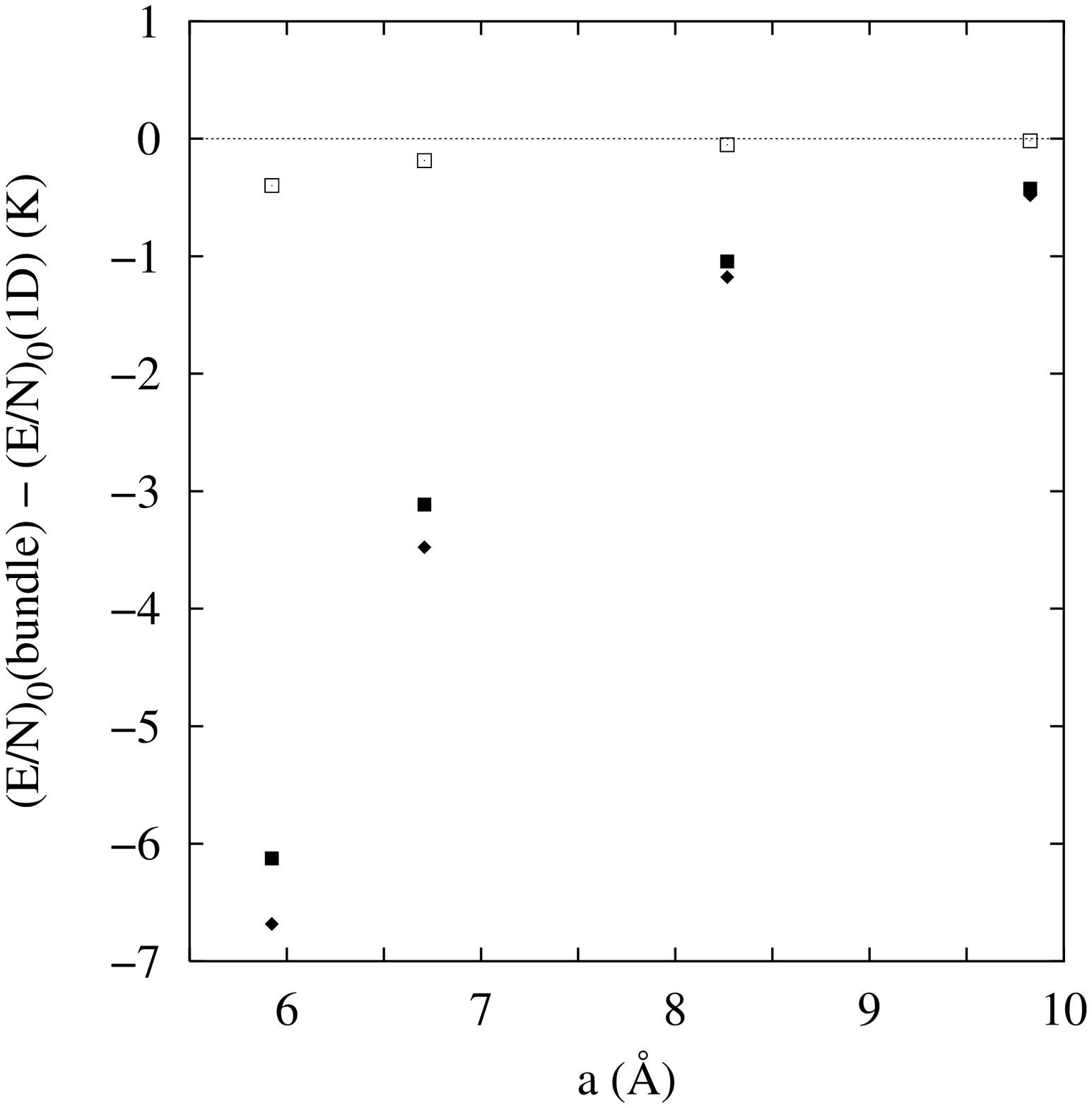}
\end{center}
\caption{Difference between the binding energies, in a bundle and in a 1D
array, as a function of the cell parameter $a$. Diamonds, filled squares,
and open squares correspond to D$_2$, H$_2$, and $^4$He, respectively.} 
\end{figure}


\begin{references}

\bibitem{iji} S. Iijima, Nature {\bf 354}, 56 (1991).

\bibitem{SDD98} R. Saito, G. Dresselhaus, and M. S. Dresselhaus, {\it Physical
Properties of Carbon Nanotubes} (Imperial College Press, London, 1998).

\bibitem{DDES98} M. S. Dresselhaus, G. Dresselhaus, P. C. Eklund,and 
R. Saito, Phys. World {\bf 11}, 33 (1998). 

\bibitem{aja} P. M. Ajayan and T. W. Ebbesen, Rep. Prog. Phys. {\bf 60},
1025 (1997).

\bibitem{dil} A. C. Dillon, K. M. Jones, T. A. Bekkedahl, C. H. Kiang, D.
S. Bethune, and M. J. Heben, Nature {\bf 386}, 377 (1997).

\bibitem{dar} F. Darkrim and D. Levesque, J. Chem. Phys. {\bf 109}, 4981
(1998).

\bibitem{stan1} G. Stan and M. W. Cole, J. Low  Temp. Phys. {\bf 110}, 539
(1998).

\bibitem{rze} M. Rzepka, P. Lamp, M. A. de la Casa-Lillo, J. Phys. Chem. B
{\bf 102}, 10894 (1998).

\bibitem{wan} Q. Wang and J. K. Johnson, J. Chem. Phys. {\bf 110}, 577
(1999).

\bibitem{tei} W. Teizer, R. B. Hallock, E. Dujardin, and T. W. Ebbesen,
Phys. Rev. Lett. {\bf 82}, 5305 (1999); Phys. Rev. Lett. {\bf 84}, 1844
(2000).

\bibitem{cole} G. Stan, M. J. Bojan, S. Curtarolo, S. M. Gatica, and M.W.
Cole,  
http://arXiv.org/abs/cond-mat/0001334.

\bibitem{kro} E. Krotscheck and M. D. Miller, Phys. Rev. B
{\bf 60}, 13038 (1999).

\bibitem{gor} M. C. Gordillo, J. Boronat, and J. Casulleras, Phys. Rev. B
{\bf 61}, R878 (2000).

\bibitem{moroni} M. Boninsegni and S. Moroni, J. Low Temp. Phys.
{\bf 118}, 1 (2000).

\bibitem{yo} M. C. Gordillo, J. Boronat, and J. Casulleras, Phys. Rev. Lett. 
{\bf 85}, 2348 (2000).

\bibitem{sin} P. Sindzingre, D. M. Ceperley, and M. L. Klein, Phys. Rev.
Lett. {\bf 67}, 1871 (1991).

\bibitem{gor2} M. C. Gordillo and D. M. Ceperley, Phys. Rev. Lett. {\bf
79}, 3010 (1997).

\bibitem{cole2} M. W. Cole, V. H. Crespi, G. Stan, C. Ebner, J. M. Hartman, 
S. Moroni, and M. Boninsegni. Phys. Rev. Lett. {\bf 84}, 3883 (2000). 

\bibitem{les} B. L. Hammond, W. A. Lester Jr., and P. J. Reynolds, {\it
Monte Carlo Methods in Ab Initio Quantum Chemistry} (World Scientific,
Singapore, 1994).

\bibitem{boro4he} J. Boronat and J. Casulleras, Phys. Rev. B {\bf
49}, 8920 (1994).

\bibitem{tipus} S. Iijima and T. Ichihashi, Nature {\bf
363}, 603 (1993).

\bibitem{sil1} I. F. Silvera and V. V. Goldman, J. Chem. Phys. {\bf 69},
4209 (1978).

\bibitem{stan2} G. Stan and M. W. Cole, J. Low Temp. Phys.  {\bf 110}, 539 (1998).

\bibitem{sie} Q. Wang, S. R. Challa, D. S. Sholl, and J. K. Johnson, 
Phys. Rev. Lett. {\bf 82}, 956 (1999). 

\bibitem{aca1} G. Baym, Phys. Rev. Lett. {\bf 17}, 952 (1966).

\bibitem{aca2} J. Boronat, A. Fabrocini, and A. Polls, J. Low Temp. Phys.
{\bf 74}, 347 (1989).

\bibitem{tersoff} J. Tersoff and R. S. Ruoff, Phys. Rev. Lett.
{\bf 73}, 676 (1994). 

\bibitem{tri1} M. K. Kostov, M. W. Cole, J. C. Lewis, P. Diep, and J. K.
Johnson, Chem. Phys. Lett. {\bf 332}, 26 (2000).

\bibitem{tri2} M. K. Kostov,J. C. Lewis, and  M. W. Cole, to appear in {\it
Many-Body Theories XXV}, edited by S. Hern\'andez;
http://arXiv.org/abs/cond-mat/0010015. 

\bibitem{axil} B. M. Axilrod and E. Teller, J. Chem. Phys. {\bf 11}, 299
(1943).

\bibitem{bruch} L. W. Bruch and I. J. McGee, J. Chem. Phys. {\bf 59}, 409
(1973).

\bibitem{com1} In Ref. \protect\onlinecite{boro4he}, the contribution 
of several three-body potentials to the
equation of state of liquid $^4$He is calculated and discussed.


\bibitem{swell2} M. M. Calbi, F. Toigo, and M. W. Cole,
http://arXiv.org/abs/cond-mat/0011509. 



\end{references}
\end{document}